\documentclass[conference]{IEEEtran}
\IEEEoverridecommandlockouts

\usepackage{epstopdf}
\def\BibTeX{{\rm B\kern-.05em{\sc i\kern-.025em b}\kern-.08em
    T\kern-.1667em\lower.7ex\hbox{E}\kern-.125emX}}

\usepackage{graphicx}

\usepackage{xcolor}

\usepackage{amssymb} 
\usepackage{amsmath}
\usepackage{amsfonts}
\usepackage{mathrsfs} 
\usepackage{tikz}

\usepackage{hyperref}

\newcommand{\doi}[1]{\href{https://doi.org/#1}{doi: #1}} 

\newcommand{\G}{\mathbb{G}}
\newcommand{\Z}{\mathbb{Z}}
\newcommand{\Scal}{\Z/\ell\Z}

\newcommand{\eg}{\emph{e.g.,}}

\newcommand{\scrE}{\mathscr{E}}
\newcommand{\scrD}{\mathscr{D}}
\newcommand{\scrR}{\mathscr{R}}
\newcommand{\scrS}{\mathscr{S}}
\newcommand{\scrK}{\mathscr{K}}

\begin{document}

\title{Polymorphic Encryption 
and Pseudonymisation of IP Network Flows\thanks{%
This work is part of the research programme VWData with project number 
400.17.605, which is (partly) financed by the Netherlands
Organisation for Scientific Research (NWO).}}


\author{
\IEEEauthorblockN{1\textsuperscript{st} Abraham Westerbaan}
\IEEEauthorblockA{\textit{Digital Security}\\
\textit{Radboud University}\\
Nijmegen, The Netherlands \\
bram@westerbaan.name}
\and 
\IEEEauthorblockN{2\textsuperscript{nd} Luuk Hendriks}
\IEEEauthorblockA{\textit{Design and Analysis of Communication Systems}\\
\textit{University of Twente}\\
Enschede, The Netherlands \\
luuk.hendriks@utwente.nl}
}

\sloppy 
\maketitle

\begin{abstract}
We describe a system, \emph{PEP3}, for storage
and retrieval
of IP flow information (equivalent to IPFIX/NetFlow)
in which the IP addresses are replaced by pseudonyms.
Every eligible party 
gets its own set of pseudonyms.
A single entity, the \emph{transcryptor},
that is composed of five independent peers,
is responsible
for the generation of, depseudonymisation of,
and translation between different sets of pseudonyms.
These operations can be performed by any three of the five peers,
preventing a single point of trust or failure.
Using homomorphic aspects of ElGamal encryption
the peers perform their operations
on encrypted --and potentially-- pseudonymised IP addresses only,
thereby never learning the (pseudonymised) IP addresses 
handled by the parties.
Moreover, using
Schnorr type proofs,
the behaviour of the peers can be verified,
without revealing the (pseudonymised) IP addresses either.
Hence the peers are central, but need not be fully trusted.
The design of our system,
while easily modified to other settings,
is tuned
to the sheer volume of data presented 
by IP flow information.
\end{abstract}


\section{Introduction}

The concept of network flow monitoring is well known and widely deployed by network operators. The
aggregation of packets passing through a certain point in the network into flow records (often based
on the 5-tuple) allows for higher-level reasoning, and moreover, it proves to be a scalable way of
measuring high quantities of network traffic.

The payload of the
packets, and thus the contents of the communication itself, is discarded in this
aggregation process: a \emph{flow} usually
only contains the number of packets and bytes that
were observed, the start and end time of that flow,
and the 5-tuple it was aggregated on.

Most routers
commonly deployed by network operators
can perform this flow aggregation, and export the
flow statistics (often using either Cisco's NetFlow \cite{rfc3954}, or IPFIX
\cite{rfc7011}, an IETF standard) towards a collector.
The collector then stores, with a certain retention time, the received flow records either on disk
or in a database, which allows the operator to query the data.

Even without the actual payload, flow measurements contain
sufficient information for a plethora of use cases, such as keeping
traffic statistics for network operators, large-scale measurement studies for
researchers, or forensic activities by security teams. 
On the flip side flow information may
reveal very sensitive information, despite
not containing the actual contents of the communication.
Consider for example the IP address of a server hosting only one
website, about a rare disease.  Or consider a server that hosts many
different videos; the
duration and size of the flow generated by a viewing of a video might betray
which video was watched.  On top of this an IP address can
constitute personal data under the GDPR.

Appreciating that flow information 
is both of great use and a potential risk,
we propose to reduce this risk
without sacrificing too much usability
by replacing the IP address in the flow data
by
pseudonyms,
and, moreover, to use a unique set of pseudonyms for every eligible \emph{party}
(storage facility, researcher, investigator, etc., further explained in
\autoref{sec:parties}).  
This might be achieved by (deterministically and
symmetrically) encrypting IP addresses destined for a certain party by a secret
key unique to---but not shared with---that party.
We hold that the \emph{transcryptor},
the entity that keeps these secret keys,
and is thus responsible for generating and translating the pseudonyms,
\begin{enumerate}
\item
should not learn the IP addresses it processes for the parties;
\item
should not have a single point of failure; and
\item
should be verifiable.
\end{enumerate}
In this paper we describe \emph{PEP3}, a system
that fulfils these requirements by:\footnote{%
    The name ``PEP3'' is a contraction of ``PEP''
    the precursor to this system, \cite{pep},
    and ``P3'', our project's designation within
    the VWData research programme.}
\begin{enumerate}
\item
having the transcryptor deal with encrypted pseudonyms
only,
and 
leveraging homomorphic aspects of ElGamal\cite{elgamal} encryption,
        noted earlier in~\cite{pep} (and arguably~\cite{pre,camenisch2015}),
to perform the required operations
on these encrypted pseudonyms;
\item
breaking the central secret keys held by the transcryptor
into ten shares each, and dividing those shares
over five independent \emph{peers}---which together form the transcryptor---in
such a way
that every triple of peers, but no pair, together has all ten shares; and
\item
verifying the honesty of the peers
    by (occasionally retroactively) requiring
        a Schnorr type proof (see \autoref{S2:verification})  
        for the performed operation.
\end{enumerate}
We envision that an organisation wishing to
run PEP3 would collaborate with multiple (up to five) independent
and possibly geographically diverse hosting providers
to run the peers,
and would arrange by contract
that the shares are not disclosed,
even to the network operator itself.

In the design of any complex system
there are a lot of knobs to turn,
and dials to watch.
We do not pretend that our system is optimal,
nor that we have defined what optimal should mean in this context.
We have, however,
preferred simplicity and speed
over cleverness and additional features.
Where possible,
we have chosen tried and trusted
cryptographic systems (curve25519, ElGamal encryption,
Schnorr type proofs)
over  exciting new techniques
(such as pairing based cryptography).
We have also carefully avoided
the need for peers to store any additional information
after a setup phase,
reducing the need for locking,
and dangers of filling volumes.

\paragraph{Contributions}
PEP3 builds on the PEP system\cite{pep} designed to store
medical research data.
Our system differs in several ways from PEP
(see \autoref{S:related-work}), 
most notably in that we divide
the three different roles (transcryptor, access manager, key server) in PEP
over five peers. 
The general idea of splitting 
a global secret into shares,
and dividing it over several peers,
so that only a subset of them is needed to perform the action,
is well-known\cite{shamir}
and fundamental to fields
such as threshold cryptography and
secure multiparty computation,
as is the use of Schnorr-type proofs in this context.
Our contribution lies primarily
in finding a combination
that fits our application.
Nonetheless,
the derivation scheme discussed in~\ref{SS:derivation}
and the ``lizard'' encoding scheme from~\ref{S2:lizard}
appear to be technical novelties.
We believe that the security, privacy, verifiability, and especially 
the simplicity 
offered
by our combination
may certainly find application
to other situations as well.

\paragraph{Demonstrator}
An interactive demonstration 
of the most important features of PEP3 can
be found at \url{https://vwdata-p3.github.io/demo.html}.
Its source code is
available at \url{https://github.com/vwdata-p3/webdemo}.

\paragraph{This paper is organised as follows}
In \autoref{S:transcryptor}
we describe the basic functioning of the transcryptor as
a single entity,
and then built it from five peers in \autoref{S:peers}.
In \autoref{S:further} we address final points by discussing technical problems,
describe how this work fits in with the GDPR and related academic works,
and spend some words on the performance of our prototype.
Finally, we conclude the paper in \autoref{S:conclusion}.

\section{Transcryptor}
\label{S:transcryptor}
\subsection{Polymorphic Encryption and Pseudonymisation}
\label{S2:pep}
\paragraph{The group}
We base PEP3 on curve25519,
an elliptic curve
introduced by Daniel Bernstein in~\cite{curve25519},
and named after the prime $p:=2^{255}-19$,
because it is both fast,
and has stood up to the scrutiny caused
by its popularity.
Using Mike Hamburg's decaf-technique\cite{decaf}
curve25519 gives rise 
to the ristretto255\cite{ristretto255,ristretto255website} group~$\G$,
a special way to represent
the cyclic group $\Z/\ell \Z$ of integers modulo
the prime
\begin{equation*}
    \ell\ :=\ 2^{252}\,+\,
    {\scriptstyle27,742,317,777,372,353,535,851,937,790,883,648,493}.
\end{equation*}
Although $\Z/\ell \Z$ and~$\G$ are isomorphic as groups,
the number~$1$ of~$\Z/\ell \Z$
being send to a specific non-zero \emph{base point}~$B$,
there is no known efficient algorithm
to find given an element~$A$ from~$\G$
the unique number~$n$ of~$\Z/\ell\Z$ with~$nB=A$.
In other words,
the \emph{discrete log problem} is difficult in~$\G$.
We also assume it to be hard
to solve the more difficult
\emph{decisional Diffie--Hellman problem}
(see \cite{ddhp})
in~$\G$,
that is,
to determine whether a triplet $(A,M,N)$
of elements of~$\G$
is a so-called \emph{Diffie--Hellman triplet},
that is,
whether writing $(A,M,N)\equiv(aB,mB,nB)$
we have $am=n$.
We will see that such hardness assumptions can be used
to show that cryptosystems 
based on~$\G$ are resistant to
several simple attacks.
The actual security provided by these cryptosystems
is, however, much more difficult to capture formally,
see Section~3 ``Security'' of~\cite{curve25519}, introducing
curve25519
for a detailed discussion.

\paragraph{IP addresses}
For now we will represent an IP address by a
non-zero element~$A$
of the group~$\G$.
We will show in \autoref{S2:lizard}
how to encode 128~bit IP addresses
(thus supporting both IPv6 and IPv4)
as elements of~$\G$.

\paragraph{ElGamal Encryption}
For the encryption of IP addresses
we use the following scheme
based on~\cite{elgamal}.
For any scalar $r\in \Scal$
the triple
\begin{equation*}
(\,rB,\,M+rsB,\,sB\,)
\end{equation*}
represents the encryption
of a message~$M\in\G$
for the \emph{private key} $s\in \Scal$.
The \emph{public key} associated with~$s$
is the element~$sB$ of~$\G$.
In general,
a \emph{cyphertext} is a triple $(\beta,\gamma,\tau)$
of elements of~$\G$,
where~$\beta$ is  the \emph{blinding},
$\gamma$ is the \emph{core},
and~$\tau$ is the \emph{target}.\footnote{%
We include the target~$\tau$ in the cyphertext
so that we can define the rerandomisation operation, $\scrR_r$,
below.}
To encrypt a message~$M\in\G$
for a public key~$\tau\in \G$,
one picks a random scalar $r\in \Scal$, and computes
\begin{equation}
\label{eq:encryption}
\scrE(M,\tau,r)\ :=\ (\,rB,\,M+r\tau,\,\tau\,).
\end{equation}
To decrypt a cyphertext $(\beta,\gamma,\tau)$
using a private key~$s\in \Scal$ one computes
\begin{equation}
    \label{eq:decryption}
    \scrD(\,(\beta,\gamma,\tau),\,s\,)\ :=\ \gamma - s\beta.
\end{equation}
Note that
$\scrD(\scrE(M,\tau,r),s)\ =\ M+r(\tau-sB)$
equals~$M$ when~$\tau=sB$.
On the other hand,
if~$\tau\neq sB$,
and~$r$ is chosen randomly,
$\scrD(\scrE(M,\tau,r),s)$ could be any element of~$\G$.

Note also that a triple
$(\beta,\gamma,\tau)$
is the cyphertext of some message~$M$
if and only if $(\beta,\tau,\gamma-M)$
is a Diffie--Hellman triplet.
Conversely,
a triplet $(A,M,N)\in \G^3$ is a Diffie--Hellman triplet
iff $(A,N,M)=\scrE(0,M,r)$ for some $r\in \Z/\ell\Z$.
So deciding whether a triple $(\beta,\gamma,\tau)$
is the cyphertext for some message~$M$ (without additional information
such as the secret key)
is thus just as difficult as the decisional Diffie--Hellman problem.
Decrypting a cyphertext without additional information such
as a guess for the plaintext
might be even harder.

\paragraph{Pseudonyms}
A \emph{pseudonym} for an IP address represented by a 
non-zero group element
$A\in\G$
is simply $nA$,
where $n\in\Scal$ is a scalar called
the \emph{pseudonym key}.
Note that depending on the pseudonym key,
the pseudonym for~$A$ could
be any element of~$\G$.

In PEP3, the transcryptor
keeps track of an \emph{encryption key} $s_P$ 
and a \emph{pseudonym key} $n_P$,
both non-zero scalars from~$\Scal$,
for every party~$P$. The encryption key~$s_P$
is shared with~$P$, while the transcryptor keeps~$n_P$ to herself.
The \emph{pseudonym for~$P$} of an IP address~$A\in \G$
is then $n_P A$,
and an \emph{encrypted pseudonym for~$P$} of~$A$
is a triple of the form
$(rB, n_PA+rs_PB,s_PB)\equiv \scrE(n_PA,s_PB,r)$
for some scalar~$r\in\Scal$.

\paragraph{Translation}
The transcryptor has the following three elementary
operations on cyphertexts 
at her disposal,
where $s,n,r\in \Scal$, cf.~\cite{pep}.
\begin{alignat*}{3}
    \text{\emph{rekeying:}}&&\qquad \scrK_s(\beta,\gamma,\tau)
    \ &:= \ (\,s^{-1}\beta,\, \gamma,\, s\tau\,)\\
    \text{\emph{reshuffling:}}&& 
    \scrS_n(\beta,\gamma,\tau)
    \ &:= \ (\,n\beta,\, n\gamma,\, \tau\,)\\
    \text{\emph{rerandomisation:}}&& 
    \scrR_r(\beta,\gamma,\tau)
    \ &:= \ (\,\beta+rB, \,\gamma+r\tau,\, \tau\,)
\end{alignat*}
To translate
an encrypted pseudonym for party~$P$
to an encrypted pseudonym for party~$Q$,
the transcryptor
applies 
\begin{equation*}
\scrK_{\smash{s_Qs_P^{-1}}}\,\scrS_{\smash{n_Qn_P^{-1}}}\,
\scrR_{r'},
\end{equation*}
where~$r'$ is some random scalar from~$\Scal$.
Note that $\scrK_{\smash{s_Qs_P^{-1}}}$ changes the target
of the encryption, sending
$\scrE(M,s_PB,r)$ to  $\scrE(M,s_QB,s_Ps_Q^{-1}r)$,
while $\scrS_{\smash{n_Qn_P^{-1}}}$ changes
the target of pseudonyms,
sending 
$\scrE(n_PM,\tau ,r)$ to  $\scrE(n_QM,\tau,n_Qn_P^{-1}r)$.

The purpose of~$\scrR_{r'}$ is more technical,
and threefold.
To begin, it prevents spoofing of the target in the cyphertext:
if the triple $(rB, M+rsB, s'B)$,
which pretends to be a cyphertext intended for~$s'B$,
but is actually decryptable by~$sB$,
is rerandomised,
the result
$(\,(r+r')B,\, M+rsB+r's'B,\,s'B\,)$
does not reveal~$M$ to
someone not knowing~$s'$.
It also prevents a party~$P$ from obtaining
the unencrypted pseudonym for~$Q$ of an IP address~$A$
by sending $(0,n_PA, s_PB)$
to the transcryptor for translation from~$P$ to~$Q$.
Finally, it makes the translation operation non-deterministic, reducing
the risk of linkability.

\paragraph{(De)pseudonymisation}
To translate a for party~$P$ encrypted IP address
to a for party~$Q$ encrypted pseudonym,
the transcryptor 
applies $
    \scrK_{\smash{s_Qs_P^{-1}}}\,\scrS_{\smash{n_Q}}\,\scrR_{r'}$,
where~$r'$ is some random scalar.
Depseudonymisation is performed similarly.

\paragraph{Polymorphism and homomorphism}
The (ElGamal) encryption scheme we use is `polymorphic' in the sense
that a message encrypted for one party can be rekeyed
to be decryptable by another party (without the need for
intermediate decryption). 
Pseudonymisation is polymorphic in a similar sense.

The fact that the translation between pseudonyms
can be performed on cyphertext makes
the encryption `homomorphic' with respect to this translation
operation.
At this point 
the transcryptor,
knowing the secret keys of the parties,
can sidestep the polymorphism and homomorphism
by first decrypting, then translating, and finally encrypting again.
However,
when the transcryptor is split into five peers in the next section,
this trick is no longer possible,
and the advantage of the polymorphic and homomorphic aspects
of the encryption become clear.

\subsection{Parties}\label{sec:parties}
Before we explain the way the transcryptor is built
(from five peers),
we will sketch how we intend her services be used
by the different parties.

\paragraph{Metering and storage}
The two most basic parties
to PEP3 
are the \emph{metering process (MP)} generating flow records,
and the \emph{storage facility (SF)} storing the flow records.
Recall that 
both parties get their own encryption keys $s_{\mathrm{MP}}$
and $s_{\mathrm{SF}}$ from the transcryptor, respectively.
After the metering process
has produced a batch 
of flow records
aggregated from packets going over the network,
it encrypts the associated IP addresses
in these flow records
using its own encryption 
key, $s_{\mathrm{MP}}$,
and sends them to the transcryptor for translation
to encrypted pseudonyms for~$s_\mathrm{SF}$,
which the transcryptor returns to the metering process.
The metering process replaces the IP addresses in the flow records
by these encrypted pseudonyms,
and sends them along to the storage facility.
Note that the metering process does not learn
the pseudonyms for
the storage facility, since they are returned by the transcryptor
to the metering process
encrypted for the storage facility's key,~$s_{\mathrm{SF}}$.
The storage facility,
having received and decrypted the encrypted pseudonyms
in the flow records, 
stores the pseudonymised flow records in its database.

\paragraph{Retrieval}
A party wishing to consult the records
held by the storage facility
may form a query in terms of their own set of pseudonyms,
and then replace their pseudonyms
by corresponding encrypted pseudonyms
for the storage facility obtained from the transcryptor.
Having received and performed the query,
the storage facility returns the result,
but only after having encrypted the pseudonyms
(from the storage facility's set)
with its encryption key, $s_{\mathrm{SF}}$.
Having received the flow record with encrypted IP addresses,
the querying party consults the transcryptor again,
this time to translate the encrypted
pseudonyms from the storage facility's set
to its own set.

\paragraph{Queries}
Clearly not every type of query should be allowed
by the storage facility
lest it runs the risk of revealing (information about) its pseudonyms.
This could not only happen directly
via a comparison between a pseudonym and a plain string,
but also by allowing, for example,
an \textsc{order by}
on a pseudonymised column.
We think a  practical solution would be to 
select a very minimal subset of SQL
such that given the information
about which columns and parameters are pseudonyms
(and which are just plain values)
the storage facility
can easily annotate each expression in the query
with either ``pseudonym'' or ``plain value''.
Instead of listing which operations on pseudonyms are inadmissible,
the storage facility 
should instead keep a list of which operations and expressions involving
pseudonyms \emph{are} admissible.
For example:
comparing a pseudonym with another pseudonym using $==$ is admissible,
and applying \textsc{count} to a pseudonym is admissible,
and \textsc{select}ing a pseudonym is admissible,
and nothing more.

\paragraph{Authorisation}
To prevent free translation between pseudonyms
and IP addresses
(defeating the pseudonymisation)
a specific permit (signed by some predetermined certification authority)
 could be required by the transcryptor for a party
wishing to perform a translation or (de)pseudonymisation.
For example,
the metering process only could be given a personal permit
to pseudonymise into storage facility pseudonyms.
A party wishing to retrieve records
from the storage facility, such as a researcher,
 would need a permit to translate pseudonyms from its own set to the set
of the storage facility (and back).
Note that if
such a researcher was to collude with the metering process,
they could link IP addresses with their own pseudonyms.

\paragraph{Depseudonymisation warrant}
A party should never be given a blanket permit
for depseudonymisation.
Instead we envision that, say, an investigator
would obtain a permit to depseudonymise a specific
(encrypted)
pseudonym from the relevant authority, after having
presenting sufficient proof to warrant this.

\section{Five Peers}
\label{S:peers}
\subsection{Ten Shares}
In PEP3, the transcryptor is split into five peers, named A, B, C, D, and E.
As a general rule,
three out of five peers should be able to act
as transcryptor.
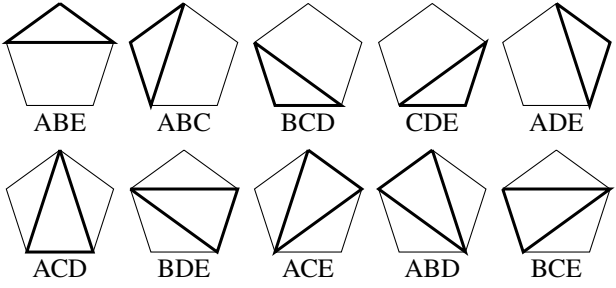
\begin{figure}
\begin{center}
\begin{tikzpicture}
    \coordinate (A) at (0.0,0.75);
\coordinate (B) at (-0.7132923872213651,0.2317627457812106);
\coordinate (C) at (-0.44083893921935485,-0.6067627457812106);
\coordinate (D) at (0.44083893921935485,-0.6067627457812106);
\coordinate (E) at (0.7132923872213651,0.2317627457812106);
\draw  (A) -- (B) -- (C) -- (D) -- (E) -- (A);
\draw[line width=.4mm] (A) -- (B) -- (E) -- (A);
\node () at (0.0,-0.8250000000000001) { ABE };
\coordinate (A) at (1.6500000000000001,0.75);
\coordinate (B) at (0.936707612778635,0.2317627457812106);
\coordinate (C) at (1.2091610607806453,-0.6067627457812106);
\coordinate (D) at (2.090838939219355,-0.6067627457812106);
\coordinate (E) at (2.363292387221365,0.2317627457812106);
\draw  (A) -- (B) -- (C) -- (D) -- (E) -- (A);
\draw[line width=.4mm] (A) -- (B) -- (C) -- (A);
\node () at (1.6500000000000001,-0.8250000000000001) { ABC };
\coordinate (A) at (3.3000000000000003,0.75);
\coordinate (B) at (2.586707612778635,0.2317627457812106);
\coordinate (C) at (2.8591610607806452,-0.6067627457812106);
\coordinate (D) at (3.7408389392193553,-0.6067627457812106);
\coordinate (E) at (4.013292387221365,0.2317627457812106);
\draw  (A) -- (B) -- (C) -- (D) -- (E) -- (A);
\draw[line width=.4mm] (B) -- (C) -- (D) -- (B);
\node () at (3.3000000000000003,-0.8250000000000001) { BCD };
\coordinate (A) at (4.949999999999999,0.75);
\coordinate (B) at (4.236707612778634,0.2317627457812106);
\coordinate (C) at (4.509161060780644,-0.6067627457812106);
\coordinate (D) at (5.390838939219354,-0.6067627457812106);
\coordinate (E) at (5.6632923872213645,0.2317627457812106);
\draw  (A) -- (B) -- (C) -- (D) -- (E) -- (A);
\draw[line width=.4mm] (C) -- (D) -- (E) -- (C);
\node () at (4.949999999999999,-0.8250000000000001) { CDE };
\coordinate (A) at (6.6000000000000005,0.75);
\coordinate (B) at (5.886707612778635,0.2317627457812106);
\coordinate (C) at (6.1591610607806455,-0.6067627457812106);
\coordinate (D) at (7.0408389392193556,-0.6067627457812106);
\coordinate (E) at (7.313292387221366,0.2317627457812106);
\draw  (A) -- (B) -- (C) -- (D) -- (E) -- (A);
\draw[line width=.4mm] (A) -- (D) -- (E) -- (A);
\node () at (6.6000000000000005,-0.8250000000000001) { ADE };
\coordinate (A) at (0.0,-1.2000000000000002);
\coordinate (B) at (-0.7132923872213651,-1.7182372542187896);
\coordinate (C) at (-0.44083893921935485,-2.556762745781211);
\coordinate (D) at (0.44083893921935485,-2.556762745781211);
\coordinate (E) at (0.7132923872213651,-1.7182372542187896);
\draw  (A) -- (B) -- (C) -- (D) -- (E) -- (A);
\draw[line width=.4mm] (A) -- (C) -- (D) -- (A);
\node () at (0.0,-2.7750000000000004) { ACD };
\coordinate (A) at (1.6500000000000001,-1.2000000000000002);
\coordinate (B) at (0.936707612778635,-1.7182372542187896);
\coordinate (C) at (1.2091610607806453,-2.556762745781211);
\coordinate (D) at (2.090838939219355,-2.556762745781211);
\coordinate (E) at (2.363292387221365,-1.7182372542187896);
\draw  (A) -- (B) -- (C) -- (D) -- (E) -- (A);
\draw[line width=.4mm] (B) -- (D) -- (E) -- (B);
\node () at (1.6500000000000001,-2.7750000000000004) { BDE };
\coordinate (A) at (3.3000000000000003,-1.2000000000000002);
\coordinate (B) at (2.586707612778635,-1.7182372542187896);
\coordinate (C) at (2.8591610607806452,-2.556762745781211);
\coordinate (D) at (3.7408389392193553,-2.556762745781211);
\coordinate (E) at (4.013292387221365,-1.7182372542187896);
\draw  (A) -- (B) -- (C) -- (D) -- (E) -- (A);
\draw[line width=.4mm] (A) -- (C) -- (E) -- (A);
\node () at (3.3000000000000003,-2.7750000000000004) { ACE };
\coordinate (A) at (4.949999999999999,-1.2000000000000002);
\coordinate (B) at (4.236707612778634,-1.7182372542187896);
\coordinate (C) at (4.509161060780644,-2.556762745781211);
\coordinate (D) at (5.390838939219354,-2.556762745781211);
\coordinate (E) at (5.6632923872213645,-1.7182372542187896);
\draw  (A) -- (B) -- (C) -- (D) -- (E) -- (A);
\draw[line width=.4mm] (A) -- (B) -- (D) -- (A);
\node () at (4.949999999999999,-2.7750000000000004) { ABD };
\coordinate (A) at (6.6000000000000005,-1.2000000000000002);
\coordinate (B) at (5.886707612778635,-1.7182372542187896);
\coordinate (C) at (6.1591610607806455,-2.556762745781211);
\coordinate (D) at (7.0408389392193556,-2.556762745781211);
\coordinate (E) at (7.313292387221366,-1.7182372542187896);
\draw  (A) -- (B) -- (C) -- (D) -- (E) -- (A);
\draw[line width=.4mm] (B) -- (C) -- (E) -- (B);
\node () at (6.6000000000000005,-2.7750000000000004) { BCE };
\end{tikzpicture}
\end{center}
\caption{The ten triples
    of peers, visualised as triangles.}
\end{figure}
To this end,
each pseudonym key~$n_P$ for a party~$P$,
is defined to be a product
\begin{equation*}
n_P\ \equiv\ 
    n_P^{\textrm{ABE}}\,
    n_P^{\textrm{ABC}}\,
    n_P^{\textrm{BCD}}\,
    n_P^{\textrm{CDE}}\,
    n_P^{\textrm{ADE}}\,
    n_P^{\textrm{ACD}}\,
    n_P^{\textrm{BDE}}\,
    n_P^{\textrm{ACE}}\,
    n_P^{\textrm{ABD}}\,
    n_P^{\textrm{BCE}}
\end{equation*}
of $10\equiv\binom{5}{3}$
\emph{shares},
one for each triple of peers.
Of course,
the share $n_P^{\text{ABE}}$ is 
given to the peers~A, B, and~E,
and so on.
Note that no two peers (such as~D and~E) have access to all shares
(D and~E do not have the share of~ABC.)
However, every triple does have access to all shares,
because any two triples drawn from 
five peers must have at least one peer in common.
The encryption key~$s_P$ for a party~$P$
is split similarly into ten shares.
For brevity's sake,
let 
\begin{alignat*}{3}
    \mathcal{T}\ :=\ 
    \{\,\text{ABE},\,&\text{ABC},\,
    \text{BCD},\,\text{CDE},\,\text{ADE},\,&\\
    &\text{ACD},\,\text{BDE},\,\text{ACE},\,
    \text{ABD},\,\text{BCE}\,\}
\end{alignat*} 
denote the set of all ten triples of peers.

\paragraph{Translation}
An encrypted pseudonym for a party~$P$
can be translated 
to an encrypted pseudonym for party~$Q$
by applying the operations
\begin{equation}
    \label{eq:peerswisetranslation}
    \scrK_{\smash{s_Q^T(s^T_P)^{-1}}}\,
\scrS_{\smash{n^T_Q(n^T_P)^{-1}}}\,
\scrR_{r^T},
\end{equation}
where~$r^T$ is a random scalar, and~$T$ ranges over~$\mathcal{T}$, in sequence.
The order in which these operations
are performed
does affect the (random component of the) cyphertext,
but not the resulting pseudonym (if the input was valid).
Naturally, any of the three peers in the triple~$T$ can perform
the operation in~\eqref{eq:peerswisetranslation}.

A translation can also be performed by three operations instead
of ten, as follows. Choose three peers,
say A, C, and~D, and
split the triples among them,
by, say, 
\begin{alignat*}{3}
    \mathcal{T}_A \,&:=\, \{\, \text{ABE},\,\text{ABC},\,\text{ADE},\,
    \text{ACD},\,\text{ACE},\,\text{ABD}\,\}\\
    \mathcal{T}_C \,&:=\, \{\,\text{BCD},\,\text{CDE},\,\text{BCE}\,\}\\
    \mathcal{T}_D\,&:=\,\{\, \text{BDE} \,\}.
\end{alignat*}
Then define, for each~$X\in\{A,C,D\}$,
\begin{equation*}
    n_P^X\ :=\  \prod_{T\in \mathcal{T}_X} n_P^T
    \qquad\text{and}\qquad
    s_P^X\ :=\  \prod_{T\in \mathcal{T}_X} s_P^T,
\end{equation*}
and have~$A$, $C$, and~$D$ perform the operations
\begin{equation*}
    \scrK_{\smash{s_Q^X(s^X_P)^{-1}}}\,
\scrS_{\smash{n^T_Q(n^X_P)^{-1}}}\,
\scrR_{r^X},
\end{equation*}
for all three $X\in \{A,C,D\}$,
in sequence,
on the encrypted pseudonym for~$P$.

(De)pseudonymisation can be performed by three
peers in a similar fashion.
\paragraph{Alternative constellations}
Our choice to divide the secrets of the transcryptor
over the triples drawn from five peers is to some extend arbitrary.
We could instead have chosen a system
where the secrets are, for example, shared among pairs drawn from three peers
(which is not resistant against collusion of two peers.)
Another option is 
to break the symmetry between peers
by 
giving a particularly important peer
 its own share, forcing its involvement in any
operation of the transcryptor.

\subsection{Verification}
\label{S2:verification}
Note that if one peer is offline, the transcryptor still functions.
Nevertheless,
a single peer
can presently  disrupt the system
in another way,
by producing erroneous results,
possibly without being detected.
One might argue
that it is possible to prevent this by having multiple peers perform the same
operation,
and compare the results.
This comparison is, however, complicated
by the random component in the encryption.
We propose a different method of verifying 
the peers' operations, namely by having the peers
attach non-interactive\cite{fiatshamir}
Schnorr type\cite{schnorr} proofs of correctness
to their results.
To keeps things simple
we create these proof
from the following basic building block.

\subsubsection{Certified Diffie--Hellman triplets}
Recall that it is considered infeasible
in general
to determine whether a triplet
$(A,M,N)$ of group elements of~$\G$
is a \emph{Diffie--Hellman triplet},
that is,
whether writing $(A,M,N)\equiv (aB,mB,nB)$
we have~$am=n$.
If the scalar~$a$ is known,
however, the matter is easily settled by checking whether
$aM=N$.
We will describe a method by which a \emph{prover} knowing~$a$
can prove to a \emph{verifier} that~$(A,M,N)$ is a Diffie--Hellman triplet,
without revealing~$a$,
using two group elements~$R_M$, $R_B$, and one scalar~$s$.
We will say that $(A,M,N)$ is \emph{certified} by~$(R_M,R_B,s)$.
\paragraph{Creating the proof $(R_M,R_B,s)$}
The prover picks a random scalar~$r\in\Scal$,
and defines $R_B:=rB$, $R_M:=rM$,
and~$s:=r+ha$, where
$h:=\mathrm{Hash}(A,M,N,R_M,R_B)$
for some appropriate predetermined hash function $\mathrm{Hash}\colon
\G^5\rightarrow \Scal$.
\paragraph{To verify a proof~$(R_M,R_B,s)$}
for the alleged Diffie--Hellman triplet~$(A,M,N)$,
the verifier computes $h:=\mathrm{Hash}(A,M,N,R_M,R_B)$,
and checks whether
\begin{equation}
\label{eq:znp}
sB\,=\,R_B+hA\quad\text{and}\quad sM\,=\,R_M+hN.
\end{equation}
\paragraph{Infeasibility of fraud}
To deceive the verifier into believing
a triplet $(A,M,N)$ is a Diffie--Hellman triplet
verified by~$(R_M,R_B,s)$
a deceiver needs to solve the two equations in~\eqref{eq:znp}.
Since the value of the hash~$h$
changes erratically depending on
its inputs,
it might as well be any
scalar as far as the deceiver is concerned.
Solving~\eqref{eq:znp} can thus be considered a game,
in which the deceiver
first chooses
$A$, $M$, $N$, $R_M$, $R_B$,
then gets a random~$h$ as challenge,
and must finally pick~$s$ such that
the two equations in~\eqref{eq:znp} hold.
Having chosen 
$A$, $M$, $N$, $R_M$, $R_B$,
the deceiver 
has a winning strategy
if and only if she has a function $s\colon \Scal\to\Scal$
with, for all~$h\in \Scal$,
\begin{equation}
    \label{eq:ws}
    s(h)B\,=\,R_B+hA\quad\text{and}\quad s(h)M\,=\,R_M+hN.
\end{equation}
Taking~$h=0$,
we see that
\begin{equation*}
s(0)B\,=\, R_B\quad\text{and}\quad
s(0)M\,=\, R_M.
\end{equation*}
Substituting this result back into~\eqref{eq:ws}
gives, after rearrangement,
for all  $h\neq 0$,
\begin{equation*}
    \frac{s(h)-s(0)}{h}\, B\,=\,A\quad\text{and}\quad 
    \frac{s(h)-s(0)}{h}\,M\,=\,N.
\end{equation*}
Hence $(A,M,N)$ is indeed a Diffie--Hellman triplet.
\paragraph{Acknowledgement}
The certified Diffie--Hellman triplets
are essentially the same as the non-interactive
version of the EQ-composition of Schnorr's protocol
in Figure~5.7 of~\cite{schoenmakers}.
\subsubsection{Certifying $\scrK_s \scrS_n \scrR_r$}
Such certified Diffie--Hellman triplets
can be used
by a peer wishing to show to a party
(via a non-interactive Schnorr type proof)
that a triple $(\beta',\gamma',\tau')$
is the result of performing
the operation $\scrK_s \scrS_n \scrR_r$
on a triple  $(\beta,\gamma,\tau)$,
that is, that
\begin{alignat*}{3}
    (\beta',\gamma',\tau')\ &= \ 
    (\,ns^{-1}(\beta + rB),\, n(\gamma+r\tau),\,s\tau\,)\\
    \ &\equiv\   \scrK_s\scrS_n\scrR_r(\beta,\gamma,\tau).
\end{alignat*}
Indeed, the peer would need only certify
the  five Diffie--Hellman triplets
\begin{alignat*}{3}
    & (\ ns^{-1}B,\ \beta+rB,\ \beta'\ ),\quad
    (\ nB,\ \gamma+r\tau,\ \gamma'\ ),
    \quad
    (sB, \tau, \tau'),
    \\
    &\qquad(\,sB, \,ns^{-1}B, \,nB\,),\quad\text{and}\quad
    (rB, \tau, r\tau).
\end{alignat*}
It's assumed here that the party
can know what~$sB$ and~$nB$ should be.
If, as in the case of translation
of a pseudonym,
$s$ is a composite,
such as $s\equiv s^T_P (s^T_Q)^{-1}$,
then
additional proof must be provided
for this
to the peer, 
for instance
a certified Diffie--Hellman triplet
$(\,s^T_P (s^T_Q)^{-1} B,\,
s^T_Q B,\, s^T_P B\,)$ in our example.

\subsubsection{Random sampling}
Having peers attach proofs to all their operations
comes at the cost of more than doubling
the number of (computationally expensive) scalar multiplications required.
This problem can be addressed by having the parties
not always request a proof of the peers, but randomly with a probability,
\eg{} $1\%$.
The request for a proof
should be put to the peer as a follow-up question, after
the operation's result has already been received by the party,
lest the peer knows when to behave properly, and when it need not.

To produce such a proof after the fact,
the peer either needs to remember the random scalars it used,
or better,
hand the random scalar to the peer (in a for the peer's eyes only 
encrypted package)
to be returned to the peer upon a request for proof.
It should, of course, be possible for a party
to resubmit its request for a proof to a peer
in case of an unexpected connection loss.
The encrypted package should, of course,
be tied to the operation (perhaps by including a hash of the original
operation in the encrypted package) to prevent 
the peer from being tricked into using the random scalar 
 in unrelated operations.

Note that a dishonest peer might be willing to risk
detection by random
sampling if its objective would be to
disrupt only a single operation
involving a specific IP address.
However, the peer can not target a specific IP address,
even if it knew its cyphertext, 
due to the rerandomisation.
At best, the peer can try to target
a specific IP address
using side-channels such as the time,
and size of the request,
forcing it, perhaps, to cast a larger net,
and increase its risk of detection.

\subsubsection{Authorisation (for depseudonymisation)}
To check whether a party is authorised
to have a certain operation performed,
a peer can often simply demand a permit,
and check it.
There is, however,
an interesting complication when the permit pertains to a specific
encrypted pseudonym, say, to depseudonymise it.
Surely, the first peer contacted
by the party can check the permit.
The second peer,
however,
being handed the partially depseudonymised encrypted pseudonym returned
by the first peer, has no way of telling whether
this partial result is related to the encrypted
pseudonym mentioned in the permit.
This problem is solved by requesting the first peer to attach
a proof to its result, which the party
can pass  along to the second peer.
Continuing in a similar fashion,
we end up with a chain of partial results and attached proofs,
that starts with the encrypted pseudonym mentioned in the warrant,
and ends with the associated encrypted plain IP address.

\subsection{Derivation of pseudonym and encryption keys}
\label{SS:derivation}
A triple~$T$ of peers
derives the pseudonym key~$n^T_P$
and encryption key~$s^T_P$ for a party~$P$
from a \emph{master pseudonym key} $n^T$,
and a \emph{master encryption key} $s^T$, respectively.
We thereby circumvent the troubles of having
to generate, store and synchronise
keys $s^T_P$ and~$n^T_P$ for every new party~$P$, on demand.
The keys are derived as follows:
assuming that each party~$P$ has some unique identifier~$\mathrm{id}_P$
from some set~$\mathcal{I}$,
and given a hash function $H\colon \mathcal{I}\to\Z/(\ell-1)\Z$,
we set
\begin{equation*}
    n^T_P\ := \ (n^T)^{H(\mathrm{id}_P)}
    \qquad\text{and}\qquad
    s^T_P\ :=\ (s^T)^{H(\mathrm{id}_P)}.
\end{equation*}
We derive the keys in this particular manner
in order to make it possible
for a peer to  give proof that $n_P^TB$
was derived from~$n^TB$,
using the~253 group elements
\begin{equation*}
n^TB,\  (n^T)^2B,\ (n^T)^4B,\  \dotsc,\ 
(n^T)^{2^{252}}B,
\end{equation*}
by, writing 
$H(\mathrm{id}_P)=\sum_{k=1}^{n} 2^{i_k}$
where
$0\leq i_1<i_2<\dotsb <i_n\leq 252$,
certifying the Diffie--Hellman triplets
\begin{alignat*}{3}
    (\quad (n^T)^{2^{i_1}}B,\quad
    &(n^T)^{2^{i_2}}B,\quad 
    (n^T)^{2^{i_1}+2^{i_2}}B \quad )\\
    (\quad  (n^T)^{2^{i_1}+2^{i_2}}B,\quad
    &(n^T)^{2^{i_3}}B,\quad 
    (n^T)^{2^{i_1}+2^{i_2}+2^{i_3}}B \quad ) \\
    &\vdots \\
    (\quad  (n^T)^{2^{i_1}+\dotsb+2^{i_{n-1}}}B,\quad
    &(n^T)^{2^{i_n}}B,\quad 
    n^T_PB \quad ).
\end{alignat*}
Such a proof for $n^T_PB$ is needed by peers and parties
wishing to check a proof (from \autoref{S2:verification})
in which~$n_P^TB$ appears.
Any party~$Q$ should be able to request such a proof for~$n_P^TB$
from a peer in the triple~$T$.
In particular,
the party~$P$ can pass along a proof of $n_P^TB$
to a peer
not in~$T$ needing proof of~$n_P^TB$
to verify, for example, a depseudonymisation request.
In this way, the peer does not need to contact the other peers.

Regarding the security of this derivation scheme:
we conjecture that recovering~$n^T$
from the group elements $(n^T)^{2^0}B,\,\dotsc,\,(n^T)^{2^{252}}B$ 
is essentially as difficult as computing the discrete log
for one of~253 random group elements.

\subsection{Setup and enrolment}
We assume that the peers and parties
can authenticate one another
and communicate securely, \eg{} by using TLS and certificates.

\paragraph{Setup}
To start PEP3
each triple~$T$ of peers needs
to decide on secrets~$n^T$ and~$s^T$,
and the public parts $n^TB$, $(n^T)^2B$, \ldots and $s^TB$, $(s^T)^2B$,
\ldots
need to be shared with the other peers.
The secret $n^{\text{ABC}}$
might be generated, for example,
by first having each of the pairs AB, BC, AC
use Diffie--Hellman key exchange
to decide on secrets $n^{\text{AB}}$,
$n^{\text{BC}}$, $n^{\text{AC}}$,
and then define
$n^{\text{ABC}}:= n^{\text{AB}}\,n^{\text{BC}}\,n^{\text{AC}}$.
Of course, B and C would both need to transmit $n^{\text{BC}}$ to~A,
and~A should check the missives agree,
and so on.
At this stage, any peer can anonymously disrupt the system by
sending incorrect key material around.
It would be preferable that dishonest
action of a peer
during setup would be identifiable by the other peers,
and we see devising a scheme providing such safeguard as a possible improvement to our system.
A perhaps simpler solution
is to
have each pair of peers compare
the public parts they obtained, and abort when 
any inconsistencies are found.

\paragraph{Enrolment}
The situation for adding a party~$P$ to the system,
so-called \emph{enrolment},
does not have the problem described in the setup phase.
The party~$P$ simply requests
the public keys
$n^TB$, $(n^T)^2B$, \ldots{} and $s^TB$, $(s^T)^2B$, \ldots{}
from every peer,
and the secret $s_P^T$ 
from every
peer in the triple~$T$,
with proof for $s_P^TB$ from $s^TB$, $(s^T)^2B$, \ldots{}.
If two peers in a triple send incorrect values for $s_P^T$,
then if the other three peers are honest,
$P$ can detect the correct value for~$s_P^T$,
and thus which peers were dishonest,
by following the majority's claim for the value 
of  $s^TB,\,(s^T)^2B,\,\dotsc$.

\section{Final points}
\label{S:further}
\subsection{Encoding IP addresses}
\label{S2:lizard}
One technical problem we encountered
when using the ristretto255 group~$\G$ was
the lack of a direct way to encode
a 128 bit piece of data~$w$
(such as an IPv6 or IPv4 address)
as an element~$\underline{w}$ of~$\G$ in such a way that
the data~$w$ can cheaply be recovered from~$\underline{w}$.
Such an encoding is useful
for encrypting~$w$ using the ElGamal scheme described 
in \autoref{S2:pep}.

The other direction presents no problem: there is a canonical and 
reversible
way to encode an element of~$\G$ as a 32-byte string,
but only $\ell / 2^{256} \approx 6.25\%$ of all 32-bytes
strings are a valid encoding
of an element from~$\G$.
So what is usually done 
(circumventing the need to encode a message as group element
before encrypting it)
is to pick  a random group element
and use its 32-byte encoding to encrypt the
message \emph{symmetrically}.

This solution is not viable for our system,
because  rerandomisation and reshuffling cannot be applied
to the symmetric cyphertext.
Instead we would like to use
\emph{elligator 2}{\cite{elligator}},
which does give
a reversible map $\mathsf{ell2}\colon 
\Z/p\Z\longrightarrow \G$,
but each element of~$\G$ can have up to~$16$ preimages
under~$\mathsf{ell2}$.

Since~$\mathsf{ell2}(x)=\mathsf{ell2}(-x)$
we can discard half the preimages
by considering only the elements of~$\Z/p\Z$
whose minimal positive representative is even.

Thus the map
$\mathsf{ell2'}\colon
    \{0,1\}^{253} \to \G$
    defined by
\begin{equation*}
    \mathsf{ell2'}(b_1\dotsb b_{253}) \ :=\  \mathsf{ell2}(\ b_1 2^1+ b_2 2^2 + \dotsb 
    + b_{253}2^{253}\  )
\end{equation*}
is reversible,
and the preimage $\mathsf{ell2'}^{-1}(A)$
of an element~$A\in \G$
has at most~8 elements.
Now,
define $\mathsf{lizard}\colon \{0,1\}^{128}\to  \G$
by
\footnote{%
    Implementations of (a trivial variation on) lizard
can be found on
\url{https://github.com/vwdata-p3/webdemo/blob/master/ed25519.py}
and \url{https://github.com/bwesterb/go-ristretto/blob/master/ristretto.go}.}
\begin{equation*}
\mathsf{lizard}(b_1\dotsb b_{128})
\ :=\  \mathsf{ell2'}(b_1\dotsb b_{128} h_1 \dotsb h_{125}),
\end{equation*}
where $h_1\dotsb h_{125}$ are the first 125 bits
of the SHA-256 hash of~$b_1\dotsb b_{128}$.
Then~$\mathsf{lizard}$ is easily computable, and reversible,
and, the preimage~$\mathsf{lizard}(A)$
of an element~$A\in \G$ almost always
contains at most~1 element.
Indeed, assuming that the bits
of a word~$w$ in the preimage  $\mathsf{ell2'}(A)$
are distributed randomly, the chance that the last
125 bits of such a word match the first 125 bits of the SHA-256 of~$w$
should be $\frac{1}{2^{125}}$.
Thus the chance that given $w\in \{0,1\}^{128}$
the preimage~$\mathsf{lizard}^{-1}(\mathsf{lizard}(w))$
contains only~$w$ is at least $(1-\frac{1}{2^{125}})^7$.
So even if $10^{10}$ computers
would apply $\mathsf{lizard}$ to~$10^{10}$  unique IP addresses 
per second for 300 years ($\approx 10^{10}$ seconds),
all  $\approx 2^{100}$
IP addresses will map to a group element with a unique preimage
with probability of at least
$(1-\frac{1}{2^{125}})^{7\cdot 2^{100}}
\geq 1-7\cdot 2^{100} \cdot \frac{1}{2^{125}}
\geq 1-\frac{1}{2^{21}}\geq \frac{999,999}{1,000,000}$.

\subsection{GDPR and pseudonymisation}
Some remarks regarding the effects of the GDPR
on our proposal are in order.

\paragraph{IP addresses as personal data}
According to the GDPR\cite{gdpr},
personal data is any information 
that can identify a natural person, either directly 
or indirectly, see article~4(1) (of the GDPR).
This means that not only the full name of a natural person,
but also just an IP address he or she used may constitute personal data.
In fact, IP addresses are explicitly mentioned as potential identifiers
in recital~30 (of the GDPR). Moreover,
the European Court of Justice
has decided that a dynamic IP address
used by a natural person to visit a website
constitutes personal data for the website operator provided
the ISP has additional data needed to identify the user by the dynamic
IP address, and the provider of the website has the legal means
to obtain this additional data\cite{dynamic-ip}.
It is thus advisable for a network
operator to treat any IP address as potential personal data.
\paragraph{Consent}
Perhaps contrary to popular belief processing of personal data does
not necessarily require consent of the data subject.
Consent is just one of six potential grounds
for lawful processing provided in article~6.1.
One of the other grounds is that the processing 
is necessary for a legitimate interest of the data controller.
In fact,
recital~49 mentions explicitly (but with qualifications)
that processing of personal data 
for the purposes of network and information security 
constitutes a legitimate interest.
It is not clear,
however, whether
this ground would cover
processing of flow data 
for research,
but it is not ruled out (in the context of clinical trails)
in point~14 of an opinion, \cite{legitimate}, 
of the European Data Protection Board.
Moreover, the Dutch research network operator
SURFnet considers requests
from researchers
for using some of its flow records 
for research, albeit on a case-by-case basis,
and
under strict conditions, see~\cite{surf}.

\paragraph{Pseudonymisation}
Pseudonymised personal data
is  still personal data,
according to recital~26.
Pseudonymisation is thus not a tool
to circumvent the processing of 
personal data, and the associated legal restrictions.
Instead,
pseudonymisation is a technique
that according to article~25.1 must be considered by
any data controller to meet its obligation
to implement data-protection principles such as data minimisation,
article~5.1(c).
It is not unthinkable
that if flow data \emph{can} be stored and used in pseudonymised form, 
it therefore \emph{must} be,
under the GDPR.

\subsection{Related work}
\label{S:related-work}
\subsubsection{PEP}
As already mentioned,
PEP3 is based on PEP,
see \url{https://pep.cs.ru.nl} and~\cite{pep},
which applies similar techniques
to store personal medical data
encrypted and pseudonymised.
One of the main selling points of PEP
is that the data subjects can control
which parties get access to their medical data.
PEP3, on the other hand, is more oriented towards helping network operators
fulfil their data protection obligations.

One important feature of PEP---not needed and thus not included
in PEP3---is the ability
to store (medical) data (such as MRI scans) 
in polymorphically encrypted form. 
Indeed, if we were to store, say, the source and destination ports
in encrypted form \emph{within the database of the storage facility},\footnote{In spite of that,
or even exactly because of it,
the database should be stored on an encrypted disk,
and flow records should be transmitted through encrypted channels.}
we could no longer use them in queries, 
hampering the usability of the flow records.

While PEP and PEP3 are closely related, conceptually PEP is more intricate.
The core functions of PEP are performed
by three entities: the transcryptor,
the access manager, 
and the key server.
In PEP an encryption key $s_P$ for a party~$P$
is of the form~$s_P \equiv k^{\text{AM}}_P k^{\text{T}}_P x$,
where~$k^{\text{AM}}_P$
and $k^{\text{T}}_P$ are secrets
that depend on~$P$, and are
known only by the access manager, and transcryptor, respectively,
while~$x$ is a secret known only to the key server and does not depend on~$P$.

To enrol in PEP, a party~$P$ contacts the key server and access manager, 
but not the transcryptor; the transcryptor is contacted 
by the access manager on the party's behalf.
The key server does not send~$x$ to~$P$ directly,
because it is important that~$x$ is kept a secret.
For the same reason the
access manager can not relay $k_P^{\text{AM}}k_P^{\text{T}}$
to the party~$P$ for enrolment,
because with $k_P^{\text{AM}}k_P^{\text{T}}$ and~$s_P$ 
the party~$P$ could compute~$x$.
Therefore the clever scheme depicted in Figure~2.3 of~\cite{pep}
is needed to enrol a party.
From the PEP3 viewpoint
one can think of~$x$ as the encryption key of a ``generic party'' $G$,
so $x=s_G$ and $k^\text{AM}_G=k^\text{T}_G=1$.
The reason that it is important that~$x\equiv s_G$
is kept a secret is that
when party makes a request to PEP
involving a data subject identifier,
this identifier is encrypted not for the party's
encryption key~$s_P$,
but instead for~$G$'s encryption key $s_G\equiv x$.
The resulting cyphertext is called a \emph{polymorphic pseudonym},
an important concept in PEP, avoided in PEP3.

In PEP, the access manager acts as the gatekeeper:
all queries intended for the storage facility
are put directly to the access manager.
This makes a more fine-grained access control possible
in PEP than in PEP3, where the peers do not learn which queries they 
facilitate.
This is the general trend: PEP3 sacrifices features of PEP
for the sake of simplicity.

The main improvement of PEP3 over PEP is the splitting
of the secrets into shares, and the addition of Schnorr type proofs,
preventing PEP3 from malfunctioning (after setup)
even if two of the five peers misbehave.
\subsubsection{Other related work}
The rekeying of ElGamal cyphertext
has appeared before
under the name \emph{atomic proxy re-encryption}, in~\cite{pre}.

Camenisch and Lehmann\cite{camenisch2015,camenisch2017} 
propose a pseudonym system
that appears rather similar to ours from a high-level viewpoint,
involving a \emph{converter} ($\approx$transcryptor),
\emph{server (\emph{$\approx$party}) specific pseudonyms},
and exploiting  homomorphic aspects of ElGamal as well.
Their system is, however, much more advanced,
employing, for example, pairing based cryptography,
and being verified by formal security models.
Their focus is currently\cite{camenisch2017}
on making the system user-auditable,
having as use case governmental databases in mind;
our focus is on making a robust and fast transcryptor.
Under the hood their converter functions quite differently from 
our transcryptor:
the converter consists of a single entity (instead of five peers),
generates pseudonyms randomly (instead of deterministically deriving them
from secrets, as PEP3 does,)
and needs to keep records on all the pseudonyms it previously generated.
The last point might be problematic when
trying to apply their system to our use case
of storing internet flow data, due
to the large number of IP addresses.

\section{Prototype, and its performance}
For simplicity,
we have built our prototype using Python
and gRPC\footnote{\url{https://grpc.github.io}},
and only use C (via cffi\footnote{\url{https://cffi.readthedocs.io/en/latest/}})
for the most important cryptographic operations.

The flowrate our prototype can handle,
while perhaps the most important performance characteristic,
is also rather difficult to pin down,
as it depends on the type of traffic:
100~Mbit of conference calls poses less of a challenge
than 100~Mbit of DNS requests and answers, as the latter results in more individual flows. But, if
those DNS requests all go to the same resolver and thus the same IP address, the caching mechanism
will have a higher cache hit rate, which is beneficial again in terms of the how many flows the system
can process in a certain time frame. The feasibility of running the PEP3 system online and keeping
up with processing the actual traffic therefore depends on the nature the traffic on that very
network.

What we can say with confidence is that throughput is limited primarily by
the number of unique IP addresses that
appear in the flows,
since each one of those IP addresses needs to be encrypted
by the metering process,
then translated with the help of at least three peers,
and finally decrypted by the storage facility.
Indeed, these three cryptographic operations,
\eqref{eq:encryption}, \eqref{eq:decryption},
and~\eqref{eq:peerswisetranslation},
alone,
account for more than $75\%$ of process time of our prototype.
The remaining $25\%$ is finely divided over
matters such as
networking, TLS, the database engine, scheduling,
random numbers generation, and so on.

The time the three main operations take
is in turn primarily determined by the number
of scalar multiplications involved:
each operation
of the form~$s\cdot M$
where $s\in \Z/\ell\Z$ and $M\in \mathbb{G}$,
takes about 125~$\mu{}s$
on our 2~GHz i7-4750HQ processor.
Scaling the basepoint, $B$,
is significantly cheaper, costing $45\,\mu{}s$. 
Encryption, \eqref{eq:encryption},
takes one general scalar multiplication ($r\tau$),
and one basepoint scaling ($rB$),
while decryption~\eqref{eq:decryption}
takes only one scalar multiplication $s\beta$.
In general a translation,
\eqref{eq:peerswisetranslation},
requires one basepoint and four general scalar multiplications,
but since in our setting the target is fixed (being the storage facility),
we need one fewer scalar multiplication.

Hence, processing one IP address should take
at least $1.6\,ms$
(based on 11~scalar multiplications,
and 4~base scalings,)
but since more than just scaling is involved
such as (un)packing points and scalars,
we see in benchmarks a cost of $2.3\,ms$ per IP address
for the three main operations,
and $3\,ms$ in total,
as our prototype processes
$20\,000$~unique IP addresses per minute
 on our quad core.

\section{Conclusion}
\label{S:conclusion}
We have described PEP3, 
a system
for pseudonymising IP flow data
built on curve25519 via the ristretto255 group~$\G$.
An important feature of PEP3 is
a robust transcryptor
(consisting of five peers)
that functions even when two peers act dishonestly.
Moreover, the peers do not learn the pseudonyms
they process,
and the the peers' actions can be verified.
We also showed how to map IP addresses to~$\G$
by a for-all-practical-purposes reversible method
(using elligator2).
Concerning future research, PEP3 would benefit greatly
from an improved way to generate its secrets
that cannot be disrupted by an anonymous peer.
Moreover,
it would be interesting to see if the flow-based intrusion detection
system SSHCure\cite{sshcure} could run against PEP3.

On the more practical side,
the next order of business for us
is to load test our internal prototype
of PEP3 by connecting it to a router generating flow data. 
We expect this experiment leads to several adjustments
to PEP3.

\bibliographystyle{IEEEtran}
\bibliography{IEEEabrv,main}

\end{document}